\documentclass[twocolumn]{article}
\usepackage[margin=0.75in,columnsep=0.25in]{geometry}
\usepackage{flushend}
\PassOptionsToPackage{hyphens}{url}
\usepackage{times}
\usepackage{titlesec}
\usepackage{hyperref}
\usepackage{graphicx}
\usepackage{float}
\usepackage{amsmath}
\usepackage{amssymb}
\usepackage{booktabs}
\usepackage{array}
\usepackage{caption}
\usepackage{fancyhdr}
\usepackage{xcolor}
\usepackage{cleveref}
\crefname{table}{Table}{Tables}

\titleformat{\section}{\large\bfseries}{\thesection}{1em}{}
\titleformat{\subsection}{\normalsize\bfseries}{\thesubsection}{1em}{}

\hypersetup{
    pdftitle={Exploited Before It Was Fixed? Vulnerability Timelines
      from Open, Machine-Readable Evidence},
    pdfauthor={Cédric Bonhomme and Alexandre Dulaunoy},
    pdfsubject={Measuring exploitation-to-patch and related vulnerability
      lifecycle intervals from open evidence},
    pdfkeywords={vulnerability lifecycle, exploitation, patch, time to
      exploit, KEV, CSAF, GCVE, VTEM, days of risk},
    colorlinks=true, linkcolor=blue, citecolor=blue, urlcolor=blue
}

\title{\textbf{Exploited Before It Was Fixed?\\
Vulnerability Timelines from Open,\\
Machine-Readable Evidence}}
\author{
    Cédric Bonhomme \\
    \textit{Computer Incident Response Center Luxembourg (CIRCL)} \\
    \href{mailto:cedric.bonhomme@circl.lu}{cedric.bonhomme@circl.lu} [\href{https://openpgp.circl.lu/pks/lookup?op=get\&search=0xA1CB94DE57B7A70D}{57B7 A70D}]
    \and
    Alexandre Dulaunoy \\
    \textit{Computer Incident Response Center Luxembourg (CIRCL)} \\
    \href{mailto:alexandre.dulaunoy@circl.lu}{alexandre.dulaunoy@circl.lu} [\href{https://openpgp.circl.lu/pks/lookup?op=get\&search=0x09E2CD4944E6CBCD}{44E6 CBCD}]
}
\date{\small 17 November 2026 --- Submitted to the FIRST.org 2027 Cyber Threat Intelligence Conference, Berlin, Germany}

\begin{document}

\makeatletter
\twocolumn[
\begin{@twocolumnfalse}
\maketitle

\begin{abstract}
How long are vulnerabilities exploited before a fix is available? We compute, for each vulnerability, the interval between exploitation and patch, and lifecycle intervals adapted from the Vulnerability Timeline and Exposure Metrics framework (VTEM), from open, machine-readable evidence: CVE and CSAF records, publisher timelines (GCVE BCP-05-X-03), several Known Exploited Vulnerabilities catalogs and sightings. An evidence model selects the earliest evidence of each lifecycle event, uses only records that describe the vulnerability itself, carries the precision of every date and traces each figure to the record that produced it; it is implemented in Vulnerability-Lookup. Over the 3,758 identifiers with exploitation evidence on one instance, disclosure is dated for 98\%, patch evidence exists for 15.7\%, and the vendor-side dates VTEM requires are never published. Most exploitation dates are catalog listings, 1,041 of them on days a catalog listed or imported a backlog. Setting those and unordered same-day pairs aside, 59 vulnerabilities have patch evidence after exploitation evidence, by a median of 7.6 days, mostly from a catalog listing to a distributor's fix. Traceability made the data correctable: one collector recorded exploit publications as exploitation, 40\% of the exploitation dates of a first measurement, and one catalog dated 11\% of its listings with the vulnerability's publication time. The intervals can be computed; what limits them is the evidence publishers release, and we identify the timeline, observation and fix dates that would close the gap.
\end{abstract}
\vspace{1em}
\end{@twocolumnfalse}
]
\makeatother

\section{Introduction}

When a vulnerability is exploited, the question that matters to a
defender is whether a fix existed, and if not, for how long the
exploitation ran before one did. The question is old. Arbaugh et al.\
proposed a life-cycle model of vulnerabilities~\cite{arbaugh2000windows}; Frei et
al.\ measured the gap between exploit and patch
availability~\cite{frei2006largescale,frei2010ecosystem}; Bilge and
Dumitra\c{s} found zero-day attacks lasting 312 days on average before
disclosure~\cite{bilge2012beforeweknewit}. Each answer rests on the data
its authors could assemble (public vulnerability databases, a vendor's
telemetry, a threat intelligence team's tracking) and is reported mainly
as aggregates. Aggregates are fragile: Householder et al.\ found a median
time to exploit publication of 2 days but a mean of 91
days~\cite{householder2020exploit}. Here we report per-vulnerability
intervals with machine-readable provenance alongside the aggregates.

In 2015, the Vulnerability Timeline and Exposure Metrics framework
(VTEM)~\cite{kouns2015strangerdanger} named the dates of a
vulnerability's lifecycle and the metrics computed from them, and observed
that vendors and coordinators seldom publish those dates. Since then the
dates have started to appear in machine-readable form: CVE records carry
publication and disclosure dates~\cite{mitre_cve}, CSAF advisories carry
dated vendor fixes~\cite{csaf20}, several catalogs list known exploited
vulnerabilities with dates~\cite{cisa_kev}, sightings record observations
of exploitation and of public proofs of concept, and GCVE's BCP-05-X-03
extension lets a publisher put the whole handling timeline in the
record~\cite{gcve2026,gcve_bcp05x03}. This paper asks how far that open
evidence gets today: for how many exploited vulnerabilities the interval
between exploitation and patch can be computed, from what kind of
evidence, and with what confidence.

We make four contributions.
\begin{enumerate}
\item An \emph{evidence model} that computes, per vulnerability, the
interval between exploitation and patch and intervals adapted from those VTEM names,
from typed evidence only. It selects the earliest evidence of each event,
uses only records that state that they describe the vulnerability, carries
the precision of every date into bounds and an explicit order, and keeps
the source of every figure (\cref{tab:anchors,tab:metrics}).
\item Its \emph{reference implementation} in
Vulnerability-Lookup~\cite{vulnerability_lookup,bonhomme2024vulnlookup},
where every vulnerability page and the timeline API show the intervals
with their evidence.
\item A \emph{measurement} over the 3,758 identifiers with exploitation
evidence on one instance: coverage of each lifecycle event, the kind of
date behind each exploitation anchor, and the intervals, split by sign and
with catalog backlogs and unordered pairs set aside.
\item \emph{Corrections of the evidence} that the traceability of the
model exposed, one of which had turned 40\% of the exploitation dates
into exploit publication dates, and the publisher-side changes that would
close the remaining gaps.
\end{enumerate}

\section{Background and Related Work}
\label{sec:related}

\subsection{Vulnerability lifecycle models}

Arbaugh, Fithen and McHugh proposed a life-cycle model for system
vulnerabilities and applied it to three case
studies~\cite{arbaugh2000windows}. Frei et al.\ measured discovery,
disclosure, exploit and patch dates at scale~\cite{frei2006largescale},
and later
used about 27,000 vulnerabilities disclosed over a decade to quantify the
systematic gap between exploit and patch availability, together with a
metric for the success of responsible disclosure~\cite{frei2010ecosystem}.

The Vulnerability Timeline and Exposure Metrics framework
(VTEM)~\cite{kouns2015strangerdanger} names nine lifecycle dates
(introduced, discovered, vendor informed, vendor acknowledged, vendor
solution, disclosure, third-party solution, exploit published and solution
implemented) and derives metrics from them: vendor response time
(acknowledgement minus notification), time to patch (solution minus
notification), total time to patch (solution minus discovery), and an
organisation's time of exposure (implementation minus solution). Its
purpose is the evaluation of vendors and products and of an
organisation's own exposure, through roll-ups over the vulnerabilities of
a product or vendor. Its authors illustrated it on vulnerabilities whose
dates they knew first-hand, and observed that vendors and coordinators
seldom publish such dates.

\subsection{Measured intervals}

Bilge and Dumitra\c{s} identified 18 vulnerabilities exploited before
their disclosure in telemetry from 11 million hosts; the attacks lasted
between 19 days and 30 months, 312 days on average, and the volume of
attacks rose by up to five orders of magnitude once the vulnerabilities
were disclosed~\cite{bilge2012beforeweknewit}. Householder et al.\
studied 75,807 CVE identifiers, of which 3,164 had public exploit code:
the median time to exploit publication was 2 days but the mean 91 days,
a contrast that recurs whenever these intervals are
summarised~\cite{householder2020exploit}. Charrier and Weiner tracked
138 vulnerabilities disclosed in 2023 and exploited in the wild, 97 of
them as zero-days, and report an average time to exploit of five days;
they note that their figures are conservative because first-exploitation
dates are rarely published, or only to the month or
quarter~\cite{charrier2024tte}. Li and Paxson examined the development
of security patches at scale~\cite{li2017patches}. Bonhomme and Dulaunoy
found SARIMAX unstable on sparse, bursty sightings, with more stable
count-based forecasts in exploratory tests~\cite{bonhomme2026sightings},
motivating source-exclusion checks here.

Here we combine typed inputs from several open sources, distinguish
exploitation from exploit publication, and keep per-vulnerability
provenance to measure evidence coverage.

\subsection{Machine-readable sources}

CVE records~\cite{mitre_cve} carry publication, reservation, update and
optional public-disclosure dates. CVE and NVD references have URLs and
tags but no dates~\cite{cve_schema}: a \texttt{patch} tag does not date
a release. Version-status entries in \texttt{affected} are undated; the
optional \texttt{timeline} is free text without event types. CSAF~2.0
advisories~\cite{csaf20} carry per-vulnerability discovery and release dates
and dated remediations, including vendor fixes. GCVE~\cite{gcve2026}
lets authorities publish under their own namespaces. BCP-05-X-03~\cite{gcve_bcp05x03}
adds publisher-supplied typed handling events. The subsequent BCP-05-X-04
draft~\cite{gcve_bcp05x04} specifies aggregated events with provenance and
optional derived metrics; X-03 events can serve as inputs.
Vulnerability-Lookup~\cite{vulnerability_lookup,bonhomme2024vulnlookup}
correlates these sources with KEV catalogs and typed sightings.

\section{Evidence Model}
\label{sec:model}

The model turns every dated statement held about a vulnerability into a
small set of \emph{anchors}, from which all durations are computed. It is
implemented in Vulnerability-Lookup (pull request~\#714, merged as
\texttt{cadc9086}) and serves both the vulnerability page and the
\texttt{/api/vulnerability/\{id\}/timeline} endpoint.

\subsection{Anchors and the earliest-evidence rule}

An input becomes a calculation input only through a structured type or
field, never through a publisher's free-text label (with one exception,
a revision stating that it adds a vulnerability to an advisory,
\cref{sec:identity}): a BCP-05-X-03 event
type, a known date field at a known location of a record (the record
root, CVE metadata, the CNA container, the CSAF tracking section or a
CSAF vulnerability entry), a CSAF remediation of category
\texttt{vendor\_fix}, a typed sighting, or an eligible KEV entry. Each
input receives a \emph{role} and an \emph{evidence basis}
(\cref{tab:anchors}). A KEV entry is eligible when it belongs to the
current KEV set and its status asserts exploitation (not suspected,
disputed or withdrawn). A KEV date equal, to the second, to the
vulnerability's publication instant is not a listing date: a catalog
that fills its listing dates with the publication time states none
(\cref{sec:classification}).

\begin{table}[t]
\centering
\caption{Anchors and their inputs.}
\label{tab:anchors}
\footnotesize
\begin{tabular}{@{}l>{\raggedright\arraybackslash}p{0.62\columnwidth}@{}}
\toprule
Anchor & Inputs \\
\midrule
$V$ discovery & \texttt{discovered} events; \texttt{dateDiscovered}; CSAF \texttt{discovery\_date} \\
$R$ report & \texttt{reported}, \texttt{received}, \texttt{affected-party-notified} events; \texttt{dateReported} \\
$A$ acknowledgement & \texttt{acknowledged} events; \texttt{dateAcknowledged} \\
$D$ disclosure & \texttt{public-disclosure}, \texttt{advisory-published} events; \texttt{datePublic}, \texttt{datePublished}; CSAF release and initial release dates \\
$X$ public exploit & published proof-of-concept sightings \\
$E$ exploitation & \texttt{exploit-observed} events; \texttt{dateExploited}; exploited sightings; eligible KEV first-seen, assertion, record and catalog-addition dates \\
$P$ patch & \texttt{fix-available} events; \texttt{dateFixed}; CSAF \texttt{vendor\_fix} remediations; patched sightings \\
\bottomrule
\end{tabular}
\end{table}

An accurate input is an upper bound on the moment it documents. A catalog
lists a vulnerability after someone saw it exploited; a sighting is
reported after the observation; even a vendor's fix date can only be at
or after the fix existed. Inputs known to be inaccurate are not used as
bounds: a placeholder listing date is set aside and a batch listing is
flagged and kept out of the intervals (\cref{sec:classification}). The
anchor of a role is otherwise the \emph{earliest} of its inputs across
every record, source and evidence basis: the tightest available bound. A later observation never overrides
an earlier report merely because observations are stronger evidence; the
basis of the selected input is kept and reported instead. The origin
$O$, the start of the lifecycle-wide metrics, follows the same rule: it is
the earliest of $V$, $R$ and $D$, since a public disclosure also bounds
when the vulnerability was known, and a discovery or report date after it
cannot be the start of the lifecycle. Each selected anchor
retains its source, record, field path, original value, precision and
basis, so that every figure can be traced to the statement that produced
it.

\subsection{Record identity}
\label{sec:identity}

A vulnerability page aggregates many records: the CVE record, advisories
from vendors, distributions and national teams, and records correlated
with it. Correlations are not identity. GCVE relationships, for
instance, link records in both directions and whatever the relationship
type, including ``related'' and ``not equal''. A linked record is
therefore a calculation input only when it states that it describes the
requested vulnerability: its own identifier, its aliases, its upstream
identifier, its CVE metadata, the CVE of a CSAF vulnerability entry, or
the CVE list of a national advisory. Within a CSAF advisory covering
several vulnerabilities, only the entry for the requested one
contributes per-vulnerability dates. An advisory can also gain a
vulnerability in a later revision, and its release date then predates
that vulnerability's disclosure: when a revision states that it adds the
requested vulnerability, the date of that revision replaces, for it, the
advisory's release date. Other records remain on the
timeline but never supply an anchor.

\subsection{Precision}

Many inputs give a day, not an instant: a bare ISO date, but also
timestamps at exactly midnight UTC, which daily feeds and many CSAF
documents use to write a day. Reading them as instants would claim an
order within the day that the evidence does not establish. An input with
day precision therefore stands for any instant of its UTC day. Every
duration carries, besides its nominal value, the bounds allowed by timestamp
precision and whether its two recorded inputs can be ordered. A
same-day exploitation sighting and vendor fix yield an interval from
$-1$ to $+1$ day with the order unknown, not a measured zero. This conservative rule also widens intervals for genuine midnight events,
including scheduled releases.

\begin{table}[t]
\centering
\caption{VTEM dates~\cite{kouns2015strangerdanger} and their equivalent
in the evidence model.}
\label{tab:vtem}
\footnotesize
\begin{tabular}{@{}p{0.42\columnwidth}p{0.5\columnwidth}@{}}
\toprule
VTEM date & Evidence model \\
\midrule
Vulnerability introduced & not available \\
Discovered & $V$ discovery \\
Vendor informed & $R$ report \\
Vendor acknowledged & $A$ acknowledgement \\
Vendor solution & $P$ patch (first of any publisher) \\
Disclosure & $D$ disclosure \\
Third-party solution & not used \\
Exploit published & $X$ public exploit \\
Solution implemented & not calculated (organisation-local) \\
\midrule
--- & $E$ exploitation (observed) \\
\bottomrule
\end{tabular}
\end{table}

\subsection{Relation to VTEM}

The model is inspired by VTEM, not an implementation of it: it keeps
VTEM's dates and metric names where open evidence supports them, and
redefines each date as the earliest evidence of the event.
\Cref{tab:vtem} maps the VTEM dates onto the anchors. The vendor being
informed and the intermediary date VTEM proposed to add are both the
report $R$. $P$ is the first patch evidence of any publisher, often a
distributor's package, not necessarily the vendor's own solution. The date a flaw was introduced is not available: affected
version ranges carry versions, not dates. A third-party solution that is not a fixed release of the
product does not establish a patch. The date a fix was implemented belongs to each
organisation and is not public. Observed exploitation $E$ is not a VTEM
date: VTEM's exploit date is the publication of an exploit, which the
model keeps as the separate anchor $X$, and a proof of concept never
counts as exploitation.

\section{Metrics}
\label{sec:metrics}

Each metric has exactly one start and one end (\cref{tab:metrics}). The
three patch metrics are kept separate rather than merged into one with
fallbacks: they answer different questions (how fast a vendor fixes once
informed, how long the whole lifecycle lasts, how long the public waits),
and a metric that silently changed its start would make its values
incomparable. Only the two origin-based metrics can start at disclosure,
and they say so.

\begin{table}[t]
\centering
\caption{Durations. $N$ is the calculation time; intervals ending at $N$
are open and flagged as ongoing. Metrics marked VTEM take their name from
VTEM but are computed from the anchors.}
\label{tab:metrics}
\footnotesize
\begin{tabular}{@{}p{0.46\columnwidth}l@{}}
\toprule
Metric & Definition \\
\midrule
Vendor response time (VTEM) & $A - R$ \\
Time to patch (VTEM) & $P - R$ \\
Total time to patch (VTEM) & $P - V$ \\
Disclosure to patch & $P - D$ \\
Time unpatched & $\max(0, (P \text{ or } N) - O)$ \\
Time to exploit & $E - O$ \\
Disclosure to exploit & $E - D$ \\
Exploitation to patch & $(P \text{ or } N) - E$ \\
Disclosure to public exploit & $X - D$ \\
Public exploit to patch & $(P \text{ or } N) - X$ \\
\bottomrule
\end{tabular}
\end{table}

\paragraph{Exploitation to patch.} $(P - E)$ is the signed gap between
recorded exploitation and patch evidence: positive when patch evidence
follows exploitation evidence, negative when it precedes it. Neither sign
establishes the order of the underlying events, since both anchors may lag
them. When $P$ is
missing the interval stays open at $N$: no recorded patch does not
establish that no patch exists, and no exploitation evidence does not
establish that no exploitation occurred. Missing anchors make a metric
unavailable rather than zero.

\paragraph{Public exploit to patch.} $(P - X)$ is the gap between exploit
and patch availability studied by Frei et
al.~\cite{frei2010ecosystem}, here computed per vulnerability from the
first published proof of concept.

\paragraph{Summaries.} Because the sign of several metrics carries their
meaning, and because zero-days and n-days produce intervals of opposite
sign, a mean over all vulnerabilities mixes two populations. We report
distributions split by sign, with medians and inter-quartile ranges, and
count exact-zero intervals and pairs whose order is unknown separately, in line with the
mean--median contrast observed by Householder et al.~\cite{householder2020exploit}.

\section{Data}
\label{sec:data}

We measure on a Vulnerability-Lookup instance as of 3~November~2026
(calculation time 2026-11-03T00:00Z), with the evidence model at
Vulnerability-Lookup commit \texttt{a9c6edf2}. The instance holds the
CVE corpus (418,263 records in its CVE~5 index), the advisories of 53 CSAF
providers, national advisories (CERT-FR, CERT-Bund, NCSC-NL, CNVD, FSTEC),
OSV and GitHub advisories, its KEV catalogs and its sightings, 97\% of
the exploited and proof-of-concept ones synchronised from the public
instance (\cref{tab:sources}).

\begin{table}[t]
\centering
\caption{Sources and the anchors they can supply.}
\label{tab:sources}
\footnotesize
\begin{tabular}{@{}>{\raggedright\arraybackslash}p{0.5\columnwidth}>{\raggedright\arraybackslash}p{0.42\columnwidth}@{}}
\toprule
Source & Anchors \\
\midrule
CVE~5 records, OSV/GitHub, national advisories & $D$ \\
CSAF advisories (53 providers) & $V$, $D$, $P$ (\texttt{vendor\_fix}) \\
BCP-05-X-03 timelines & $V$, $R$, $A$, $D$, $E$, $P$ \\
KEV catalogs (6,470 exploited entries) & $E$; $P$ (ENISA \texttt{patchedSince}) \\
Exploited sightings (55,815) & $E$ \\
Proof-of-concept sightings (26,925) & $X$ \\
Patched sightings (23) & $P$ \\
\bottomrule
\end{tabular}
\end{table}

The exploited KEV entries come from Previdian (3,302 evidence records;
the catalog was formerly named KEVIntel), CISA KEV (1,766), Shadowserver's
honeypot feed (1,283), the EU CSIRTs Network catalog published by ENISA
(82), CIRCL's sinkhole (6) and two single entries. Exploited sightings include posts collected in
real time from public and private Telegram channels and classified by a
local language model; each links to its post and can be verified by hand.
Proof-of-concept sightings include the Exploit-DB entries
(\cref{sec:classification}).

\paragraph{Population.} We measure every identifier with exploitation
evidence: an eligible KEV entry or an exploited sighting. There are
3,758 (3,690 of them CVEs). Each is passed through the same builder as the
public timeline endpoint, read-only. For two identifiers the only
exploitation evidence postdates the calculation time, so 3,756 have an
exploitation anchor. The measurement script, the per-identifier output and its
metadata are published with this paper.

\section{Results}
\label{sec:results}

\subsection{Coverage}

Disclosure is almost always dated, patches rarely, and the vendor-side
dates VTEM needs never (\cref{tab:coverage}). No identifier with
exploitation evidence carries a report to its vendor or an
acknowledgement, so VTEM's time to patch and vendor response time cannot
be computed for any of them. The 362 discovery anchors are CSAF
\texttt{discovery\_date} values, all but one Red Hat's: the date the
distributor learned of the vulnerability rather than its discovery. Only
590 identifiers (15.7\%) have both an exploitation and a patch anchor;
3,168 have no patch evidence at all, 3,166 of them with an exploitation
anchor.

\begin{table}[t]
\centering
\caption{Anchor coverage over the 3,758 identifiers with exploitation
evidence.}
\label{tab:coverage}
\footnotesize
\begin{tabular}{@{}lrr@{}}
\toprule
Anchor & Identifiers & Share \\
\midrule
$E$ exploitation & 3,756 & 99.9\% \\
$D$ disclosure & 3,688 & 98.1\% \\
$X$ public exploit & 1,367 & 36.4\% \\
$P$ patch & 590 & 15.7\% \\
$V$ discovery & 362 & 9.6\% \\
$R$ report, $A$ acknowledgement & 0 & 0.0\% \\
\midrule
$E$ and $P$ & 590 & 15.7\% \\
\bottomrule
\end{tabular}
\end{table}

Patch evidence is concentrated: of the 590 patch anchors, 323 are Red Hat
and 209 Microsoft CSAF vendor fixes, 40 come from the ENISA catalog's
\texttt{patchedSince} and the rest from a handful of
sources. For most libraries the first recorded fix is a distributor's
package, not the upstream release, which carries no dated record here.

\subsection{What the exploitation dates are}

\Cref{tab:exploitation} classifies the exploitation anchors by the kind of
date they carry. Catalog listings dominate: 2,401 anchors are the date a
KEV catalog listed the vulnerability, and 87 more are sightings mirroring
such a listing. First-hand observations, the first detection by a
honeypot or sinkhole, give 875 anchors, all but two from Shadowserver.
Classified channel posts give 343.

\begin{table}[t]
\centering
\caption{Kind of date behind the exploitation anchor, over all
identifiers and over those with a patch anchor.}
\label{tab:exploitation}
\footnotesize
\begin{tabular}{@{}lrr@{}}
\toprule
Kind of date & All & With $P$ \\
\midrule
Catalog listing (KEV entry) & 2,401 & 475 \\
First-hand observation (KEV entry) & 875 & 22 \\
Classified channel post (sighting) & 343 & 81 \\
Catalog listing (sighting) & 87 & 1 \\
Other sightings & 50 & 11 \\
\midrule
Total & 3,756 & 590 \\
\bottomrule
\end{tabular}
\end{table}

Catalogs also list in batches. On 19 days a single catalog listed 25 or
more vulnerabilities: CISA KEV listed 287 on 2021-11-03, its first day,
and the import of Shadowserver's history listed 324 on 2023-10-14.
Imports into the instance do the same: on 12 days at least 25 entries of
one catalog were recorded, 2,523 of Previdian's on 2026-06-19. A record
date is an exploitation anchor only when the entry states no earlier
date, 144 times here, all on such a day. 1,041 exploitation anchors fall
on a listing or import batch day. They date the catalog, not the
exploitation, which began earlier, often years earlier.

\subsection{Exploitation to patch}

Of the 590 identifiers with both anchors, 135 have an exploitation anchor
on a batch day; we set them aside. Of the remaining 455, the order of the
two dates is unknown for 89, because at least one is known only to the day
and both fall on the same day. Patch evidence precedes the exploitation
evidence for 307 (median $-188.3$ days, IQR $-1544.4$ to $-28.4$), and
follows it for 59 (median 7.6 days, IQR 3.8 to 22.4;
\cref{tab:intervals}).

The negative side mostly records a vulnerability patched long before a
catalog listed it, not a patch that preceded the first exploitation: the
exploitation anchor is a listing for most of these pairs. The positive
side quantifies the gap between the recorded anchors. Of the 59
vulnerabilities with patch evidence after exploitation evidence, the patch
anchor follows within a day for 5, a week for 27, and 30 days for 49; for
3 it follows after more than 90 days (at most 489 days). For 50 of them the exploitation anchor is a catalog
listing (median 7.2 days), for 5 a first-hand observation (22.7 days),
for 3 a classified channel post and for one another sighting; 42 of their patch anchors are Red Hat
fixes. What the data supports is therefore narrower than ``time from
exploitation to patch'': it is mostly the time from a catalog listing to
the first recorded, frequently downstream, fix.

\begin{table}[t]
\centering
\caption{Durations over closed intervals whose order is established,
batch listings excluded (exploitation metrics). Days; median and IQR.}
\label{tab:intervals}
\footnotesize
\setlength{\tabcolsep}{4pt}
\begin{tabular}{@{}lrrr@{}}
\toprule
Metric & $n$ & Median & IQR \\
\midrule
Exploitation to patch, $>0$ & 59 & 7.6 & 3.8 -- 22.4 \\
Exploitation to patch, $<0$ & 307 & $-188.3$ & $-1544.4$ -- $-28.4$ \\
Disclosure to exploit, $>0$ & 2,389 & 298.0 & 27.6 -- 1316.8 \\
Disclosure to exploit, $<0$ & 69 & $-4.3$ & $-24.4$ -- $-1.5$ \\
Disclosure to patch, $>0$ & 365 & 13.3 & 5.1 -- 42.2 \\
Total time to patch, $>0$ & 332 & 11.4 & 4.9 -- 39.9 \\
Disclosure to public exploit, $>0$ & 1,337 & 610.0 & 75.4 -- 1593.0 \\
\bottomrule
\end{tabular}
\end{table}

\subsection{Other intervals}

Exploitation evidence follows public disclosure by a median of 298 days
when batch listings are set aside, and precedes it for 69 vulnerabilities
(median 4.3 days before). Patch evidence follows disclosure by a median
of 13.3 days (365 identifiers), coincides at the recorded timestamp for
147, and precedes it for 13, 11 of them through ENISA \texttt{patchedSince} dates. The first published proof of concept
follows disclosure by a median of 610 days. It is the first proof of
concept \emph{recorded} here, an Exploit-DB publication or a channel
post; an exploit published elsewhere earlier leaves no dated trace in this
data.

\subsection{Effect of evidence classification}
\label{sec:classification}

The figures above were preceded by four corrections of the evidence,
each visible only once every anchor could be traced to its record. In a
first measurement, on 28~October, 2,225 of 5,620 exploitation anchors
(40\%) were Exploit-DB
entries: the collector feeding them stored each entry as an
\emph{exploited} sighting, dated with the publication of the exploit,
following a taxonomy revision that had no proof-of-concept value. An
Exploit-DB entry is public exploit code, VTEM's ``exploit published''. Once
the entries were retyped as proofs of concept, the population fell from
5,620 to 3,574 identifiers, the identifiers ``exploited before
disclosure'' from 458 to 86, and the median of patch evidence after
exploitation from 39.5 to 10.6 days. Second, the first detections of
Shadowserver's honeypots tied with their listing date and were selected
as reports; recognising first-hand evidence turned 949 anchors into
observations, and flagging batch listings removed the catalog-launch dates
from the intervals. Third, 375 of Previdian's 3,302 entries (11\%) date
their listing with the CVE record's publication instant, to the second,
for vulnerabilities published from 2008 on: the catalog states no listing
date for them. Read as listings, they more than doubled the
vulnerabilities ``exploited before disclosure'' (157 instead of 69), most
by a fraction of a second, and replaced honeypot first detections by
dates years earlier. Set aside, they leave the entry's record date, the
day the instance imported the backlog, flagged as a batch listing.
Fourth, the instance had stored SUSE's copy of about 3,600 Red Hat
advisories, which SUSE republishes under Red Hat's identifiers, in place
of Red Hat's own. Restoring them made 49 patch anchors earlier and
supplied 12; none moved later. Of the 79 identifiers whose patch evidence followed
exploitation evidence in the 28~October run, 55 remain, 16 now fall on an import day
and for 8 the patch evidence now comes first, 5 of them through the restored Red Hat fixes;
the following week's data adds 4. None of the corrections changed the
arithmetic, yet each moved the figures.

\section{Threats to Validity}
\label{sec:threats}

\paragraph{Exploitation dates are upper bounds at best.} Every exploitation
anchor is the date the evidence was produced, not the date exploitation
began, and it bounds exploitation only if that date is accurate. 2,401 of 3,756 are catalog listings; only 875 are first
detections by a honeypot or sinkhole, and a honeypot sees attempts against
its own sensors, from the day it watches for them. Batch listings
are flagged and set aside, and a catalog date that is the publication
instant to the second is not read as a listing
(\cref{sec:classification}); a placeholder date in any other form would
go unnoticed. A catalog that lists a vulnerability a few
days after exploitation began is indistinguishable from one that lists it
the same day. If its anchor is accurate, exploitation began no later than it, which can only
shorten an interval that starts there; but the first recorded patch can be
later than the first fix (below), which lengthens it. The two errors do
not cancel in a known direction, so an exploitation-to-patch value is
neither a lower nor an upper bound on the true exposure.

\paragraph{Selection.} The population is what this instance's catalogs
and sightings recorded. A KEV catalog lists what its maintainer observed
and chose to publish, with its own scope; CISA's, for instance, is driven
by the needs of US federal agencies. Vulnerability statistics inherit the
selection biases of their databases~\cite{christey2013bias}: the
figures describe the evidence that is published and machine-readable, not
the population of exploited vulnerabilities.

\paragraph{Patch dates.} The patch anchor is the first fix recorded on
the instance. For most libraries it is a distributor's package (Red Hat) or a vendor advisory (Microsoft); the upstream fix usually has no
dated record, so the patch date can be months after the upstream release.
The ENISA catalog's \texttt{patchedSince} dates are statements of a CSIRT;
for 11 vulnerabilities they precede disclosure, which suggests that some
are approximate. A fix's availability says nothing about its deployment.

Patch evidence without a date is not used. Records often say that a fix
exists, and which release carries it, without saying when it was
released, for which the format has no field (\cref{sec:related}). Of the 3,168 identifiers without patch evidence, 821
(25.9\%) have a CVE or NVD reference tagged \texttt{patch}, 182 of them to
a commit or pull request. The version survey finds CNA version-status
candidates for 795 (25.1\%) and NVD range boundaries for 1,251 (39.5\%),
1,503 (47.4\%) in either. These are not verified fixes: unaffected versions
may predate the flaw, and excluded bounds need not name released
versions~\cite{cve_schema}. Dating them requires verifying the product,
branch, release and fix. Only 88 of the 795 CNA records supply a
\texttt{repo} or \texttt{collectionURL} with \texttt{packageName} for a
lookup. A tag is also not a release, and two
records from outside the population show the error in both directions:
one, whose timeline has only \texttt{fix-developed} and whose reference is
the commit, is reported unpatched although its named fixed version was
released 20 days before disclosure; in the other, the reference tagged
\texttt{patch} is a commit in no release yet, a fix no user can install. Patch coverage is therefore the coverage of dated fix evidence,
not of fixes.

\paragraph{Discovery dates.} The only discovery dates in the population
are CSAF \texttt{discovery\_date} values of distributors, which record when
the distributor learned of the vulnerability: 192 of the 362 (53\%) are
later than the public disclosure, and 18 even follow the distributor's
own fix in the same advisory. Total time to patch therefore measures a distributor's
awareness-to-fix time, not VTEM's discovery-to-patch time; such dates are
flagged, and they do not start the lifecycle-wide metrics.

\paragraph{Disclosure dates.} A disclosure date is a document's date, and
a document covering several vulnerabilities need not have disclosed each
of them on that date. Advisories gain vulnerabilities in later revisions,
industrial-control advisories in particular, which are updated for years
as the third-party components of a product gain vulnerabilities. Before
such revisions were recognised, 36 of 3,688 disclosure anchors (1\%)
were the release date of an advisory that added the
vulnerability later; recognising the revisions moves them later by a
median of 14 days and up to 723. A 2026
vulnerability was dated by a 2024 advisory, and its disclosure-to-patch
interval was 744 days instead of 21. Once the revision is recognised, the
anchor of 32 of the 36 comes from an earlier record, the CVE record for
15: the revision was not the first disclosure either. Revisions are
recognised by their wording (``Added CVE-\ldots'', ``Ajout des
identifiants\ldots'', ``\ldots{} erg\"anzt'', or the identifiers named as new); an unrecognised one
leaves the release date, which makes a disclosure date early and an
interval that starts there long. Exploitation
to patch does not depend on disclosure.

\paragraph{Classification.} Sighting types are assigned by their
collectors. One collector's choice, recording Exploit-DB entries as
exploitation, made 40\% of the exploitation anchors in our first
measurement (\cref{sec:classification}). Telegram sightings are classified
by a local language model; each can be verified against its post, but the
classifier's precision on the sightings that become anchors (343
exploitation, and the proof-of-concept ones) is unmeasured. Removing the 81
pairs anchored on a channel post from the 455 outside batch days, without
replacing their anchors, leaves the positive side almost unchanged (56,
median 7.6 days, IQR 3.7--22.1) but moves the negative median from
$-188.3$ to $-127.7$ days (229 pairs). A proof-of-concept sighting does not
establish that the code works.

\paragraph{Precision.} 439 of the 590 exploitation/patch intervals
(74.4\%) have at least one endpoint known only to the day, a bare date or
a timestamp at exactly midnight UTC. The bounds cover timestamp precision only, at most a day in either
direction; reporting delays and missing earlier evidence remain unbounded.
The order of 89 pairs is unknown, reported apart.

\paragraph{Snapshot and identity.} The measurement reads one instance on
one day. Feeds, catalogs and sightings change continuously: two
identifiers gained their exploitation evidence on the day of the
measurement, after the calculation time, and a synchronisation of the KEV
catalogs before the measurement added 595 entries, a week of listings. Linked records count only when
they state that they describe the vulnerability, which excludes records
without such fields (some national databases) and errs towards fewer,
safer inputs.

\section{Conclusion}
\label{sec:conclusion}

Open, machine-readable evidence is enough to compute, for each
vulnerability, the interval between exploitation and patch, together with
lifecycle intervals adapted from those VTEM named in 2015, and to trace every
figure to the record that produced it, with its precision and the kind of
evidence behind it. What it is not yet enough for is the claim the
interval invites. Of 3,758 identifiers with exploitation evidence, 590
have patch evidence at all; most exploitation dates are catalog listings;
the vendor-side dates VTEM needs are never published; and once catalog
backlogs and unordered same-day pairs are set aside, 59 vulnerabilities
remain whose patch evidence follows their exploitation evidence, by a
median of 7.6 days, mostly from a catalog listing to a distributor's fix.

Two results matter beyond the numbers. First, traceability is what made
the data correctable: a single collector's classification of exploit
publications as exploitation moved 40\% of the exploitation dates, and it
was found because each anchor names its source; so were a catalog's
placeholder listing dates, which had moved a quarter of the
vulnerabilities whose patch followed their exploitation. Second, the gaps are
specific and can be closed by publishers. Timelines in the records
themselves (GCVE BCP-05-X-03: reported, acknowledged, fix available,
exploitation observed) would supply the vendor-side dates; catalogs that
publish a first observation separately from their listing date would turn
listing dates, upper bounds at best, into observations; and dated upstream fixes, a
\texttt{fix-available} event rather than an undated \texttt{patch}
reference or a \texttt{fix-developed} event alone, would remove the
distributor lag from patch dates.

Future work follows from these gaps: the precision of the channel-post
classifier; dating fix evidence at import time, once a candidate fixed
version or fix commit is verified, from the first release that carries it;
repeated measurements as
publishers adopt timelines; and the roll-up by product and vendor that was
VTEM's original purpose.

\section*{Reproducibility and Artifacts}
\begin{itemize}
  \raggedright
  \item Code: Vulnerability-Lookup~\cite{vulnerability_lookup}
        (\href{https://github.com/vulnerability-lookup/vulnerability-lookup}{github.com/vulnerability-lookup/vulnerability-lookup},
        AGPL-3.0). The evidence model is
        \path{website/lib/vulnerability_timeline.py}, documented in
        \path{docs/vulnerability-timeline.md}. Pull request \#714
        (\texttt{cadc9086}) introduced it; \#753 (first-hand observations,
        batch listings, catalog patch dates), \#764 (origin), \#765 and
        \#768 (revisions adding a vulnerability), \#766 and \#767 (copied
        publication times, import days) and \#769 (Red Hat advisories in
        the SUSE feeds, with \path{tools/fix_csaf_suse_redhat_copies.py})
        refined it. Every figure is computed at \texttt{a9c6edf2}. \#752
        added the \texttt{retype\_sightings} command of
        \cref{sec:classification}.
  \item Collector: ExploitDBSighting
        (\href{https://github.com/vulnerability-lookup/ExploitDBSighting}{github.com/vulnerability-lookup/ExploitDBSighting}),
        pull request \#4 (Exploit-DB entries are proofs of concept) and
        commit \texttt{553eb0f} (unverified entries imported).
  \item Per-vulnerability evidence: every Vulnerability-Lookup instance
        serves the events and durations of a vulnerability, each with its
        source, record, field, precision and basis, at
        \path{/api/vulnerability/{id}/timeline}, for instance on
        \href{https://vulnerability.circl.lu}{vulnerability.circl.lu}.
  \item Measurement: \path{scripts/measure_coverage.py} in the paper
        repository runs, read-only, inside a Vulnerability-Lookup
        checkout, through the same builder as the endpoint above, and
        records the commit, the calculation time and the batch days.
        \path{scripts/paper_figures.py} derives every figure of
        \cref{sec:results,sec:threats} from its output, and
        \path{observation_sensitivity.py} details the source exclusions;
        \path{scripts/instance_stats.py}, \path{patch_references.py} and
        \path{fixed_versions.py} give those of \cref{sec:data} and the
        undated patch evidence.
  \item Results: \path{results/} holds the per-identifier output
        (anchors with source, record and field; durations with bounds,
        order and batch flags) of every run. Historical exports omit original
        date values, entry UUIDs and some diagnostic flags; new exports retain
        them. The figures of this paper are those of \path{coverage-2026-11-03-paper} (\texttt{a9c6edf2}, calculation
        time 2026-11-03T00:00Z) and of the instance and survey files of
        that day; the comparisons use the same data with the model before
        \#765 (\path{-pre765}). Earlier runs document the corrections of
        \cref{sec:classification}, and \path{NOTES.md} records each run and
        the decisions between them.
  \item Data: the instance itself is not redistributed. Its sources are
        public (CVE records, CSAF providers, national advisories, KEV
        catalogs) or served by the public instance (sightings, KEV
        entries). A run on another instance or another day will differ;
        the per-identifier output is what makes each figure here
        traceable.
\end{itemize}

\bibliographystyle{plain}
\bibliography{references}

\end{document}